\documentclass{article}
\usepackage{spconf}
\usepackage{amsmath}
\usepackage{amssymb}
\usepackage{graphicx}
\usepackage{cite}
\usepackage{epsf}
\usepackage{times}
\usepackage{float}
\usepackage{color}
\usepackage{wasysym}
\usepackage{subeqnarray}
\usepackage{colortbl}
\usepackage{booktabs}
\usepackage{subcaption}
\usepackage{algorithm}
\usepackage{algorithmicx}
\usepackage{algpseudocode}
\usepackage{multirow,multicol}
\usepackage{tikz}
\usepackage{amsmath}
\usepackage{bm}
\usepackage{makecell}
\graphicspath{{Figures/}}
\usepackage{dcolumn}
\usepackage{epstopdf}
\usepackage{setspace}

\ninept

\makeatletter

\renewcommand\section{%
	\@startsection{section}{1}{0pt}%
	{-5pt} 
	{5pt}   
	{\normalfont\large\bfseries}%
}

\renewcommand\subsection{%
	\@startsection{subsection}{2}{0pt}%
	{-2pt} 
	{2pt}  
	{\normalfont\normalsize\bfseries}%
}
\makeatother

\title{Forward Convolutive Prediction for Frame Online Monaural Speech Dereverberation Based on Kronecker Product Decomposition}
\name{
	\begin{tabular}{c}
	Yujie Zhu$^1$, Jilu Jin$^2$, Xueqin Luo$^2$, Wenxing Yang$^3$, \\
	Zhong-Qiu Wang$^4$, Gongping Huang$^1$, Jingdong Chen$^2$, and Jacob Benesty$^5$
	\end{tabular}
}
\address{\begin{tabular}{c}
$^1$School of Electronic Information, Wuhan University, Wuhan, Hubei, China \\
$^2$CIAIC, Northwestern Polytechnical University, Xi'an, Shaanxi 710072, China \\
$^3$
School of Oriental Pan-Vascular Devices Innovation College, University of Shanghai for \\
Science and Technology, Shanghai, 200093, China \\
$^4$
Department of Computer Science and Engineering, Southern University of \\
Science and Technology, Shenzhen, China \\
$^5$INRS-EMT, University of Quebec,  Montreal, QC H5A 1K6, Canada
\end{tabular}}
\begin{document}
%
\maketitle
\begin{abstract}
Dereverberation has long been a crucial research topic in speech processing, aiming to alleviate the adverse effects of reverberation in voice communication and speech interaction systems.
Among existing approaches, forward convolutional prediction (FCP) has recently attracted attention.
It typically employs a deep neural network to predict the direct-path signal and subsequently estimates a linear prediction filter to suppress residual reverberation. However, a major drawback of this approach is that the required linear prediction filter is often excessively long, leading to considerable computational complexity. 
To address this, our work proposes a novel FCP method based on Kronecker product (KP) decomposition, in which the long prediction filter is modeled as the KP of two much shorter filters. This decomposition significantly reduces the computational cost. An adaptive algorithm is then provided to iteratively update these shorter filters online. 
Experimental results show that, compared to conventional methods, our approach achieves competitive dereverberation performance while substantially reducing computational cost.
\end{abstract}
\section{Introduction}
\label{Sect-Intro}
In room acoustic environments, speech signals captured by microphones generally comprise not only the direct-path component but also reflections from walls, ceilings, and other surrounding surfaces, collectively referred to as reverberation~\cite{benesty2008microphone, benesty2023microphone}. Reverberation severely degrades speech intelligibility, adversely impacting both human perception and the performance of speech-related technologies~\cite{kinoshita2013reverb, kinoshita2016summary, schmid2014variational}.
Over the past few decades, numerous dereverberation techniques have been developed to alleviate these adverse effects~\cite{schwartz2016expectation, kodrasi2016joint}. 
Among them, linear prediction-based methods have been intensively studied~\cite{nakatani2019unified, kamo2022importance, nakatani2008blind, yang2018dereverberation}. They typically employ a linear prediction filter to estimate the late reverberation component, which is subsequently subtracted from the observed signal to obtain the desired signal. Notably, the weighted prediction error (WPE) algorithm has emerged as a particularly powerful approach~\cite{nakatani2010speech, ikeshita2021online}. Its online variant, the adaptive WPE (AWPE)~\cite{yoshioka2009adaptive} algorithm, has subsequently been developed and demonstrates strong potential in practice, owing to its robust dereverberation performance in dynamically changing acoustic environments.
However, the high computational complexity of WPE and AWPE presents a challenge for their practical deployment.
To alleviate this issue, recent studies have explored linear filtering-based dereverberation methods leveraging the Kronecker product (KP) decomposition~\cite{yang2021robust, yang2022bilinear, huang2022kronecker, huang2023switching}. Specifically, the long global filter to be estimated is expressed as the KP of shorter filters, which significantly reduces both the number of parameters and the computational burden.

Another issue is that, under adverse acoustic conditions involving background noise or interference sources, the performance of single-channel WPE can degrade significantly. To overcome these limitations, Wang et \emph{al.}~\cite{wang2021convolutive, wang2023unssor, wang2024usdnet} introduced the forward convolutional prediction (FCP) method, a hybrid approach that integrates deep learning with linear prediction for single-channel speech dereverberation. 
Although FCP can provide improved performance compared to WPE in single-channel dereverberation, it still demands substantial computational cost due to the high order of the filter.

In this work, we integrate the FCP method into the previously proposed KP decomposition framework and present a KP decomposition-based dereverberation approach. 
The core idea is to represent the original prediction filter as the KP of two shorter filters, thereby greatly reducing the cost of overall computational complexity. 
These two shortened filters are then iteratively updated according to the recursive least-squares (RLS) concept. In practice, both the conventional FCP considered in this paper and the proposed KP-based FCP are implemented in a frame-wise online manner. 
Experimental results demonstrate that the proposed method matches or surpasses FCP in dereverberation performance, while greatly reducing computational complexity.

\section{Signal Model and Problem Formulation}
\label{Sect-SM-PF}

Let us consider a noisy and reverberant environment, the signal received by a single microphone in the short-time Fourier transform (STFT) domain can be expressed as~\cite{wang2021convolutive}
\begin{align}
\label{Y-vect}
Y\left(t, f\right)&=X\left(t, f\right) + V\left(t, f\right) \nonumber \\
&= S\left(t, f\right) +\sum^\ell_{i = 0} H \left( i, f \right) S \left( t - i, f \right) + V\left(t, f\right) \nonumber   \\
&=S\left(t, f\right) + S_{\mathrm{r}}\left(t, f\right) + V\left(t, f\right) ,
\end{align}
where $t$ and $f$ denote the time-frame and frequency-bin indices, respectively, $X\left(t, f\right)$ is the reverberant target speech in the STFT domain, which can be conceptually divided into two parts: the direct path $S\left(t, f\right)$ and the early reflections plus late reflections i.e. $S_{\mathrm{r}}\left(t, f\right) = \sum^\ell_{i = 0} H \left( i, f \right) S \left( t - i, f \right)$, where $H \left( i, f \right)$ is the early and late parts of the room impulse response (RIR) in the STFT domain, and $Y \left( t, f \right)$ and $V \left( t, f \right)$ are the STFT-domain representations of the observed signal and background additive noise, respectively.

The goal of this work is to estimate the direct-path component $S\left(t, f\right)$ from the noisy reverberant observation $Y\left(t, f\right)$.

\section{Forward Convolutive Prediction for Speech Dereverberation}

This section provides a brief review of the FCP method~\cite{wang2021convolutive, wang2023unssor, wang2024usdnet}.
FCP first employs a deep neural network (DNN) to estimate the direct-path component, denoted by $\hat{S}_{\mathrm{nn}}\left(t, f\right)$. The residual signal $Y\left(t, f\right) - \hat{S}_{\mathrm{nn}}\left(t, f\right)$ is regarded as estimated reverberation under the assumption that $V\left(t, f\right)$ is weak. It is then estimated by forwardly filtering the DNN-estimated direct-path signal $\hat{S}_{\mathrm{nn}}\left(t, f\right)$, where the filter is obtained by solving the following problem~\cite{wang2021convolutive}:
\begin{align}
\label{FCP}
\operatorname*{argmin}_{\mathbf{g}^\prime\left(t, f\right)} \sum_{t} \frac{\left|Y\left(t, f\right) - \hat{S}_{\mathrm{nn}}\left(t, f\right) - \mathbf{g}'^H\left(t, f\right) \hat{\mathbf{s}}_{\mathrm{nn}}\left(t, f\right)  \right|^2}{\hat{\lambda}\left(t, f\right)},
\end{align}
where $\hat{\mathbf{s}}_{\mathrm{nn}}\left(t, f\right)=\left[\hat{S}_{\mathrm{nn}}(t, f), \hat{S}_{\mathrm{nn}}(t-1, f), \ldots, \hat{S}_{\mathrm{nn}}\right.$$(t-K+1, f)]^{T}$ and $\hat{\lambda} \left(t, f\right)$ is a weighting factor that adjusts the relative significance of different time–frequency (TF) units.

By absorbing $\hat{S}_{\mathrm{nn}}\left(t, f\right)$ into $\hat{\mathbf{s}}_{\mathrm{nn}}\left(t, f\right)$, (\ref{FCP}) can be reexpressed as 
\begin{align}
\label{FCP-2}
\operatorname*{argmin}_{\mathbf{g}\left(t, f\right)} \sum_{t} \frac{\left|Y\left(t, f\right) - \mathbf{g}^H\left(t, f\right) \hat{\mathbf{s}}\left(t, f\right)  \right|^2}{\hat{\lambda}\left(t, f\right)}.
\end{align}

The dereverberation result is obtained as 
\begin{align}
\label{dereverberation}
\hat{S}\left(t, f\right) &= Y\left(t, f\right) - Z\left(t, f\right),
\end{align}
where $Z\left(t, f\right) = \mathbf{g}'^H\left(t, f\right) \hat{\mathbf{s}}_{\mathrm{nn}} \left(t, f\right) = \mathbf{g}^H\left(t, f\right) \hat{\mathbf{s}}_{\mathrm{nn}}\left(t, f\right) - \hat{S}_{\mathrm{nn}}\left(t, f\right)$ is estimated reverberation.

Although the FCP method has demonstrated excellent dereverberation performance, it requires updating a long filter $\mathbf{g}^H\left(t, f\right)$ at each frequency band, leading to high computational complexity and limiting its applicability. Therefore, in the following sections, we propose a decomposition method based on the KP, which reduces computational complexity by decomposing the long filter into two shorter ones.

\section{Kronecker Product FCP for Speech Dereverberation}
\label{Sect-Method}

\begin{figure}[t!]
\centerline{\includegraphics[width=80mm]{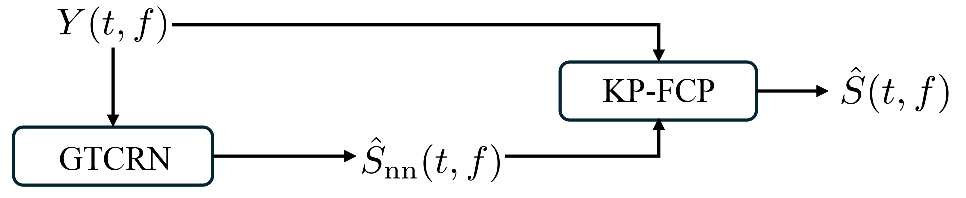}}
\caption {Proposed frame-online dereverberation system. }
\label{System}
\end{figure}

Building on the idea of FCP, our frame-online speech dereverberation system adopts a two-stage architecture, as illustrated in Fig.\ref{System}. In the first stage, a causal DNN, called grouped temporal convolutional recurrent network (GTCRN)~\cite{rong2024gtcrn}, operates in an online manner to estimate the direct-path component of the microphone observation, with its architecture briefly described in Subsection \ref{Sect-MA}. In the second stage, a linear prediction filter is applied to suppress the reverberation. 
Unlike the conventional FCP method, the proposed system factorizes the linear prediction filter into the KP of two shorter filters, as detailed in Subsection \ref{KP-FCP}, achieving preserving dereverberation performance with substantially reduced computational complexity.

\subsection{Model Architecture }\label{Sect-MA}

\begin{figure*}[t!]
\centerline{\includegraphics[width=150mm]{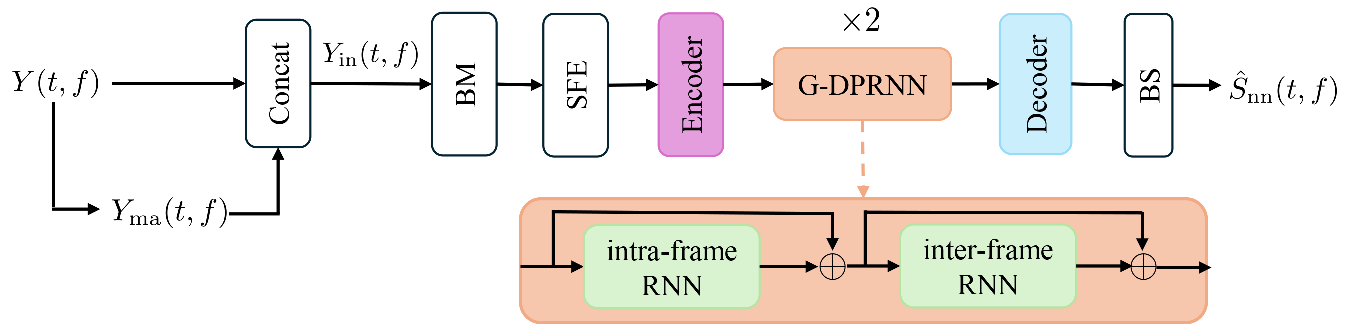}}
\caption {Architecture of the online DNN used to estimate the direct sound component. }
\label{Net}
\end{figure*}

The architecture of the online causal DNN for direct sound component estimation is shown in Fig.~\ref{Net}. For the observed signal $Y\left(t, f\right)$, it is concatenated with its magnitude spectrum $Y_\mathrm{ma}\left(t, f\right)$ to form $Y_{\mathrm{in}}\left(t, f\right)$ as the input. First, a band merging (BM) module compresses the high-frequency part of $Y_{\mathrm{in}}\left(t, f\right)$. Immediately after, the subband feature extraction (SFE) module unfolds and reshapes adjacent frequency bands to enrich inter-frequency relationships.
The encoder encodes the extracted features into a compact time–frequency (T-F) representation, which is then processed by two grouped dual-path RNN (G-DPRNN) modules for intra-frame and inter-frame modeling. Since our goal is to build a causal system, unidirectional gated recurrent units (GRUs) are used in the G-DPRNN modules for inter-frame modeling, ensuring that the network does not access information from future frames.
Finally, the decoder together with the band splitting (BS) operation predicts the direct-path spectral component $\hat{S}_{\mathrm{nn}}\left(t, f\right)$. In the implementation, the network is realized in a streaming frame-online manner, meaning its outputs are generated frame by frame.

\subsection{FCP Based on Kronecker Product Decomposition}
\label{KP-FCP}

Exploiting the low-rank property of the linear prediction filter $\mathbf{g}\left(t, f\right)$, we represent it within the KP framework as in \cite{paleologu2018linear, huang2022kronecker}:
\begin{align}
\label{g-decomp-P}
\mathbf{g}\left(t, f \right) = \sum_{p=1}^{P} \mathbf{g}_{2,p}\left(t, f \right) \otimes \mathbf{g}_{1,p}\left(t, f \right),
\end{align}
where $\otimes$ denotes the KP operator, $\mathbf{g}_{2,p}\left(t, f \right)$ and $\mathbf{g}_{1,p}\left(t, f \right)$, $p=1,2,\dots,P$  are two sets of short filters with lengths of $K_2$ and $K_1$, such that $K = K_2 K_1$ with $K_2 \le K_1$, and $P$ denotes the order of KP decomposition, which generally satisfies $P \leq \min\left(K_1, K_2\right)$ following the singular value decomposition (SVD) principle.

It can be proved that we have the two relationships given below for the KP ~\cite{harville1998matrix}:
\begin{align}
\label{g-kp-1}
\mathbf{g}\left(t, f \right)
&= \sum_{p=1}^{P} \left[ \mathbf{g}_{2,p}\left(t, f \right) \otimes \mathbf{I}_{K_1} \right] \mathbf{g}_{1,p}\left(t, f \right) \nonumber \\
&= \sum_{p=1}^{P} \left[ \mathbf{I}_{K_2} \otimes  \mathbf{g}_{1,p}\left(t, f\right)\right] \mathbf{g}_{2,p}\left(t, f\right),
\end{align}
where $\mathbf{I}_{K_1}$ and $\mathbf{I}_{K_2}$ are the $K_1 \times K_1$ and $K_2 \times K_2$ identity matrices. Expression~(\ref{g-kp-1}) can be rearranged as
\begin{align}
\label{G-2-arr}
\mathbf{g}(t, f) 
& = \underline{\mathbf{G}}_{2}\left(t, f\right)\underline{\mathbf{g}}_{1}\left(t, f\right) \nonumber\\
&= \underline{\mathbf{G}}_{1}\left(t, f\right)\underline{\mathbf{g}}_{2}\left(t, f\right),
\end{align}
where
\begin{align}
\label{g-1}
\underline{\mathbf{g}}_{1}\left(t, f\right)
&=  \left[\mathbf{g}_{1,1}^T\left(t, f\right) ~ \mathbf{g}_{1,2}^T\left(t, f\right) ~ \cdots  ~ \mathbf{g}_{1,P}^T\left(t, f\right)  \right]^T, \\
\underline{\mathbf{g}}_{2}\left(t, f\right)
&=  \left[\mathbf{g}_{2,1}^T\left(t, f\right) ~ \mathbf{g}_{2,2}^T\left(t, f\right) ~ \cdots  ~ \mathbf{g}_{2,P}^T\left(t, f\right)  \right]^T
\end{align}
are vectors of length $PK_1$ and $PK_2$, respectively,
\begin{align}
\label{G-2}
\underline{\mathbf{G}}_{2}\left(t, f\right)
&= \left[ \mathbf{G}_{2,1}\left(t, f \right)
~ \mathbf{G}_{2,2}\left(t, f \right) ~ \cdots ~ \mathbf{G}_{2,P}\left(t, f \right) \right ], \\
\underline{\mathbf{G}}_{1}\left(t, f\right)
&= \left[ \mathbf{G}_{1,1}\left(t, f \right)
~ \mathbf{G}_{1,2}\left(t, f  \right) ~ \cdots ~ \mathbf{G}_{1,P}\left(t, f \right) \right ],
\end{align}
with 
\begin{align}
\label{G-2p}
\mathbf{G}_{2,p}\left(t, f \right)
&=  \mathbf{g}_{2,p}\left(t, f\right) \otimes \mathbf{I}_{K_1}, \\
\label{G-1p}
\mathbf{G}_{1,p}\left(t, f \right)
&=  \mathbf{g}_{1,p}\left(t, f\right) \otimes \mathbf{I}_{K_2}.
\end{align}

Substituting (\ref{G-2-arr}) into (\ref{dereverberation}), the dereverberated signal can be rewritten as
\begin{align}
\label{dr-out-g1}
\hat{S}_1\left(t, f \right)
&= \hat{S}_{\mathrm{nn}}\left(t, f \right) + Y \left(t, f \right) - \underline{\mathbf{g}}_{1}^H\left(t, f \right)
\underline{\hat{\mathbf{s}}}_{2} \left(t, f \right) \\
& =\hat{S}_{\mathrm{nn}}\left(t, f \right) + Y \left(t, f \right) - \underline{\mathbf{g}}_{2}^H\left(t, f \right)
\underline{\hat{\mathbf{s}}}_{1} \left(t, f \right),
\end{align}
where
\begin{align}
\label{y_2p_define}
\underline{\hat{\mathbf{s}}}_{2} \left(t, f \right)
&=\underline{\mathbf{G}}_{2}^H\left(t, f \right) \hat{\mathbf{s}}_\mathrm{nn}\left(t, f \right), \\
\underline{\hat{\mathbf{s}}}_{1} \left(t, f \right)
&=\underline{\mathbf{G}}_{1}^H\left(t, f \right) \hat{\mathbf{s}}_\mathrm{nn}\left(t, f \right)
\end{align}
are vectors of length $PK_1$ and $PK_2$, respectively.

According to the RLS concept, a cost function can be formulated, and the two short filters can be iteratively solved following the procedure in~\cite{huang2022kronecker}. The detailed solution steps are summarized in Algorithm~\ref{Algo-KP-FCP}.
As a result, the two filters $\underline{\mathbf{g}}_{1}\left(t, f \right)$ and $\underline{\mathbf{g}}_{2}\left(t, f \right)$ are updated iteratively.
To maintain consistency with the conventional FCP algorithm \cite{wang2021convolutive}, the proposed method is referred to as the KP-FCP.

\begin{algorithm}[t!]
\footnotesize
\setstretch{1.56}
\caption{The KP-FCP Algorithm}
\label{Algo-KP-FCP}
\begin{algorithmic}[1]
\State Initialize $\underline{\mathbf{g}}_{1}\left(0, 0 \right)$, $\underline{\mathbf{g}}_{2}\left(0, 0 \right)$, $\bm{\Phi}_{\underline{\hat{\mathbf{s}}}_{2}}^{-1} \left(0, 0 \right)$, $\bm{\Phi}_{\underline{\hat{\mathbf{s}}}_{1}}^{-1} \left(0, 0 \right)$, $\alpha_1$, $\alpha_2$
\For{$t=1,2,\dots$}
\For{$f=1,2,\dots$}
\State $ \begin{aligned}
\lambda \left(t, f \right) = \left|Y\left(t, f\right) \right|^2  +  \sigma \mathrm{max} \left(\left|Y(i, j) \right| \right)^2, 
&i = 1, 2,\cdots, t,  \\
&j = 1, 2,\cdots, f
\end{aligned} $
\State $e_{1} \left(t, f \right) = Y \left(t, f \right) - \underline{\mathbf{g}}_{1}^H\left(t-1, f \right) \underline{\hat{\mathbf{s}}}_{2} \left(t, f \right)$
\State update $\bm{\Phi}_{\underline{\hat{\mathbf{s}}}_{2}}^{-1} \left(t. f \right)$ and $\underline{\mathbf{g}}_{1}\left(t, f \right)$:
\State
$\bm{\kappa}_{2} \left(t, f \right) = \frac{\bm{\Phi}_{\underline{\hat{\mathbf{s}}}_{2}}^{-1} \left(t-1, f \right) 
\underline{\hat{\mathbf{s}}}_{2} \left(t, f \right)}
{\alpha_1 \lambda \left(t, f \right) + \underline{\hat{\mathbf{s}}}_{2}^H \left(t, f \right)
\bm{\Phi}_{\underline{\hat{\mathbf{s}}}_{2}}^{-1} \left(t-1, f \right) 
\underline{\hat{\mathbf{s}}}_{2} \left(t, f \right)}$
\State   $ \begin{aligned}
\bm{\Phi}_{\underline{\hat{\mathbf{s}}}_{2}}^{-1} \left(t, f \right)
&= \frac{1}{\alpha_1} [ \bm{\Phi}_{\underline{\hat{\mathbf{s}}}_{2}}^{-1} \left(t-1, f \right) \\ 
&~~~~~~~~~~~~~~
-\bm{\kappa}_{2}\left(t, f \right) 
\underline{\hat{\mathbf{s}}}_{2}^H \left(t, f \right)
\bm{\Phi}_{\underline{\hat{\mathbf{s}}}_{2}}^{-1} \left(t-1, f \right) ]
\end{aligned} $    
\State $\underline{\mathbf{g}}_{1}\left(t, f \right) = \underline{\mathbf{g}}_{1}\left(t-1, f \right) + \bm{\kappa}_{2}\left(t, f \right) e^{*}_{1} \left(t, f \right)$

\State $e_{2} \left(t, f \right) = Y \left(t, f \right) - \underline{\mathbf{g}}_{2}^H\left(t-1, f \right) \underline{\hat{\mathbf{s}}}_{1} \left(t, f \right)$
\State update $\bm{\Phi}_{\underline{\hat{\mathbf{s}}}_{1}}^{-1} \left(t, f\right)$ and $\underline{\mathbf{g}}_{2}\left(t,f  \right)$:
\State $\bm{\kappa}_{1} \left(t, f \right) = \frac{\bm{\Phi}_{\underline{\hat{\mathbf{s}}}_{1}}^{-1} \left(t-1, f \right) \underline{\hat{\mathbf{s}}}_{1} \left(t, f \right)}
{\alpha_2 \lambda \left(t, f \right) + \underline{\hat{\mathbf{s}}}_{1}^H \left(t, f \right)
\bm{\Phi}_{\underline{\hat{\mathbf{s}}}_{1}}^{-1} \left(t-1, f \right) 
\underline{\hat{\mathbf{s}}}_{1} \left(t, f \right)}$
\State   $ \begin{aligned}
\bm{\Phi}_{\underline{\hat{\mathbf{s}}}_{1}}^{-1} \left(t, f \right)
&= \frac{1}{\alpha_2} [ \bm{\Phi}_{\underline{\hat{\mathbf{s}}}_{1}}^{-1} \left(t-1, f \right) \\ 
&~~~~~~~~~~~~~~
-\bm{\kappa}_{1}\left(t, f \right) 
\underline{\hat{\mathbf{s}}}_{1}^H \left(t, f \right)
\bm{\Phi}_{\underline{\hat{\mathbf{s}}}_{1}}^{-1} \left(t-1, f \right) ]
\end{aligned} $    
\State $\underline{\mathbf{g}}_{2}\left(t, f \right)= \underline{\mathbf{g}}_{2}\left(t-1, f \right) + \bm{\kappa}_{1}\left(t, f \right) e^*_{2}\left(t, f \right)$
\State 
$\hat{S}\left(t, f\right) = \hat{S}_{\mathrm{nn}}\left(t, f\right) + Y \left(t ,f \right) - \underline{\mathbf{g}}_{2}^H\left(t, f \right) \underline{\hat{\mathbf{s}}}_{1} \left(t, f \right).$

\EndFor
\EndFor
\end{algorithmic}
\end{algorithm}

\section{Simulations}

\subsection{Training Configuration}

The clean speech signals are taken from the VCTK dataset~\cite{yamagishi2019cstr}, whose training set includes $34,647$ utterances. All utterances are resampled to $16$~kHz.
To simulate reverberant acoustic environments, the image method~\cite{Allen1979image} are used to artificially generate RIRs.
We consider room sizes ranging from $5 \times 5 \times 3$~m$^3$ to $10 \times 10 \times 4$~m$^3$, with a microphone placed at the center of the room at a height between $1$~m and $2$~m. The distance between the sound source and the microphone varies from $0.5$~m to $2$~m.
The reverberation time $T_{60}$ is randomly generated between $0.3$~s and $0.8$~s.
The clean signals are first convolved with the corresponding simulated impulse responses to generate the microphone signals. Then, Gaussian white noise with a signal-to-noise ratio (SNR) between $20$~dB and $30$~dB is added to obtain the observed microphone signals. 
All time-domain signals are transformed into the STFT domain using a frame size of $512$ and an overlap of $75\%$. 
The AdamW optimizer is used for training, with the initial learning rate set to $5\times 10^{-4}$ and scheduled to decay by a factor of $0.98$ at each epoch.

\subsection{Algorithm Implementation}
The parameters for the FCP and KP-FCP methods are configured as follows: $K = 81$, $K_1 = 9$, and $K_2 = 9$. We initialize the inverse matrices $\bm{\Phi}_{\underline{\hat{\mathbf{s}}}_{2}} \left(t, f \right)$ and $\bm{\Phi}_{\underline{\hat{\mathbf{s}}}_{1}} \left(t, f \right)$  as identity matrices. The recursive factors $\alpha_1, \alpha_2$ are set to $0.95$, and $\delta$ is set to $0.01$.
To ensure consistency with the KP-FCP algorithm, we adapt the FCP algorithm into FCP (online). In particular, the RLS concept is employed to iteratively optimize the filter in a frame-online manner, and the network for direct-path signal estimation is replaced with GTCRN. 
The quality of speech enhancement is evaluated using the perceptual evaluation of speech quality (PESQ) and frequency-weighted segmental signal-to-noise ratio (FWSNR)~\cite{kinoshita2016summary, hu2007evaluation}. 
We show their performance improvement in terms of gain in PESQ and FWSNR, denoted, respectively, as $\Delta$PESQ and $\Delta$FWSNR~\cite{huang2022kronecker}.

\subsection{Computational Complexity Analysis}

\begin{figure}[t!]
\centerline{\includegraphics[width=95mm]{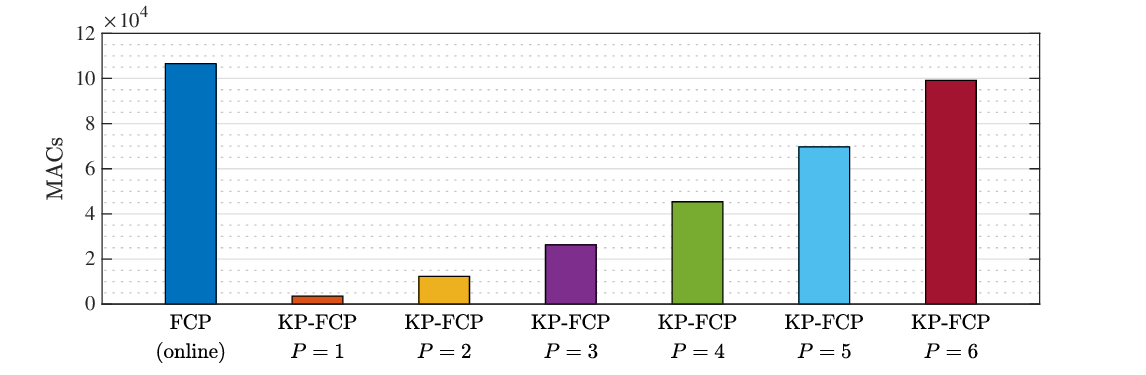}}
\caption {Computational complexity in terms of MACs of the FCP (online) and KP-FCP methods with $K = 81$, $K_1 = 9$, and $K_2 = 9$.}
\label{Fig-Sim-2}
\end{figure}

In the dereverberation system shown in Fig.~\ref{System}, the computational cost of GTCRN is about $2.1$~K multiply accumulate (MAC) per TF unit, where MAC refers to the number of multiply-accumulate operations. This cost is far lower than that of the subsequent FCP-series methods, which can even be ignored.
Therefore, the main computational burden of the system lies in the subsequent FCP-series methods.
Table~\ref{Computational Complexity} compares the computational complexity of the FCP (online) and KP-FCP algorithms. The complexity of the KP-FCP method is $\mathcal{O}(P^2K^2_{2}+P^2K^2_{1})$, whereas the FCP (online) exhibits a higher complexity of $\mathcal{O} (K_{2}^2K_{1}^2)$. Considering that $P < K_{1}$, the computational complexity of KP-FCP can be significantly lower than that of FCP (online). 
Figure~\ref{Fig-Sim-2} shows the curves of MACs of two methods as a function of $P$. 
It can be seen that when $P < 6$, the computational complexity of KP-FCP is clearly lower than that of FCP (online).

\begin{table}[!t]
\renewcommand{\arraystretch}{1.3}
\caption{Theoretical computational complexity of the FCP (online) and KP-FCP methods (note that $K =K_1 K_2$).}
\label{Computational Complexity}
\centering
{\footnotesize
\begin{tabular}{c|c}
\hline \hline
Method & MACs \\
\hline
\makecell{FCP\\(online)} & $16K ^2+20K +16$ \\
\hline
KP-FCP & $16P^2(K^2_{1}+K^2_{2})+8PK_{1}K_{2}+16PK_{1}+20PK_{2}+24$ \\
\hline \hline
\end{tabular}
}
\end{table}

\subsection{Simulation Results}

\begin{figure}[t!]
\centerline{\includegraphics[width=95mm]{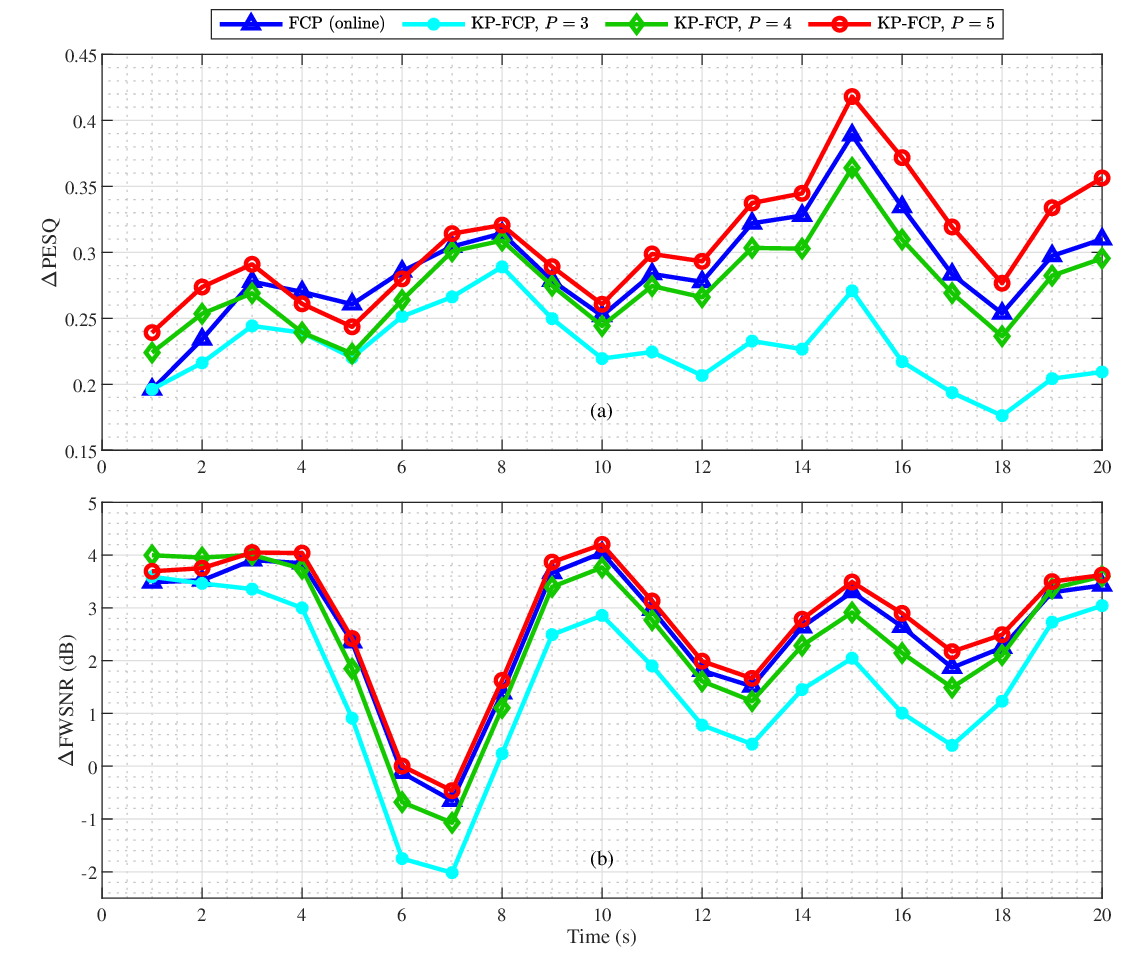}}
\caption {Improvement in PESQ, and FWSNR of the FCP (online) and KP-FCP methods: (a) $\Delta$PESQ and (b) $\Delta$FWSNR. Experimental conditions: $\mathrm{SNR} = 25$~dB, $T_{60} = 400$ ms, $K = 81$, $K_1 = 9$ and $K_2 = 9$.}
\label{Fig-Sim-1}
\end{figure}

We first processed a 20‑second simulated reverberant speech signal in a $7 \times 7\times 3$~m$^3$ room ($T_{60}=400$~ms), using the FCP (online) and KP-FCP algorithms. 
After convolving the clean signal with the RIR, white noise with an SNR of  $25$~dB is added.
The segmental performance was computed at one-second intervals, and the resulting scores were further smoothed using a three-second moving average. 
Figure~\ref{Fig-Sim-1} depicts smoothed $\Delta$PESQ and $\Delta$FWSNR results for different $P$ values. 
It can be seen that both methods obtain significant improvement in the PESQ and FWSNR. 
As $P$ increases, the performance improvement of KP-FCP becomes more significant. While its performance is marginally less than FCP (online) when $P=3$, it reaches or even exceeds FCP (online) for $P=4$  and $P=5$.

\setlength{\tabcolsep}{1.5mm}{
\begin{table}[t!]
\renewcommand{\arraystretch}{1.3}
\centering
\caption{Overall average performance of FCP (online) and KP-FCP methods.} \vskip -8pt
\label{results_dr}
\begin{tabular}{c|cc|cc|cc}
\hline \hline
~&\multicolumn{2}{c|}{$T_{60}=400$~ms}&\multicolumn{2}{c|}{$T_{60}=500$~ms}&\multicolumn{2}{c}{$T_{60}=700$~ms}\\
\cline{2-7}
~ & PESQ& \makecell{\rule{0pt}{2.4ex} FWSNR \\ (dB)} & PESQ& \makecell{ FWSNR \\ (dB)}& PESQ& \makecell{\rule{0pt}{2ex} FWSNR \\ (dB)}\\
\hline
observed&1.411&1.661& 1.344 &1.003& 1.258 &-0.075\\
\makecell{FCP\\(online)} & 1.709& 4.803 & 1.622& 4.220& 1.556 &3.777 \\
\makecell{KP-FCP\\($P=3$)}& 1.672 & 4.609& 1.590 &3.595& 1.525 &3.595 \\
\makecell{KP-FCP\\($P=4$)}& 1.764 & 5.308& 1.671 &4.346& 1.608 &4.346 \\
\makecell{KP-FCP\\($P=5$)}& \textbf{1.837}&\textbf{5.790}& \textbf{1.735} & \textbf{5.214} & \textbf{1.661} & \textbf{4.850} \\
\hline \hline
\end{tabular} \vskip -12pt
\end{table}
}

Subsequently, we evaluated these dereverberation methods on the test set of the VCTK dataset~\cite{yamagishi2019cstr}, including $872$ utterances. 
The reverberation times were set to three levels: $400$~ms, $500$~ms, and $700$~ms, with the same RIR settings as in training.
Table~\ref{results_dr} shows the overall average performance of the FCP (online) and KP-FCP.
As $P$ increases, the KP-FCP performance improves. Once $P$ reaches a high value, the performance of KP-FCP can even exceed that of FCP (online), but with increased computational complexity, necessitating a judicious choice of $P$ to achieve a favorable trade-off between dereverberation performance and efficiency in practice.

\section{Conclusions}
\label{Sect-Conclu}

In this study, we addressed the challenge of balancing dereverberation performance and computational efficiency in frame-online speech signal processing. By integrating the FCP framework with the KP decomposition, we proposed an efficient dereverberation method that models the long prediction filter as the KP of two shorter filters, substantially reducing the computational burden while preserving or improving dereverberation performance. 
Experimental evaluations under various reverberant conditions confirmed that the proposed approach achieves a favorable trade‑off between computational complexity and speech quality.

\bibliographystyle{ieeetr}
\bibliography{Bib_MABFSE}

\end{document}